\documentclass[manuscript]{acmart}
\AtBeginDocument{%
  }

\setcopyright{none}
\acmISBN{}
\copyrightyear{2026}
\acmYear{2026}

\usepackage{graphicx}
\usepackage{url}   
\usepackage{xcolor}
\usepackage{soul}
\usepackage[most]{tcolorbox}
\usepackage[dvipsnames]{xcolor}

\newif\ifredact
\redacttrue   

\newif\ifcomment
\commenttrue   
\ifcomment
    \newcommand{\missing}[2][]{\textcolor{red}{[\textbf{MISSING\ifx#1\empty\else~–~#1\fi}] ~#2}}
    \newcommand{\kel}[1]{~\sethlcolor{pink}\hl{[Kellie: #1]}}
    \newcommand{\ken}[1]{~\sethlcolor{yellow}\hl{[Kenny: #1]}}
\else
  \newcommand{\missing}[2]{}
  \newcommand{\kel}[1]{}
  \newcommand{\ken}[1]{}
\fi

\begin{document}
\settopmatter{printacmref=false}
\renewcommand\footnotetextcopyrightpermission[1]{}

\title[Redistributing Voice and Responsibility: AI in Relationship-Centred Care]{Redistributing Voice and Responsibility: AI in Relationship-Centred Care}

\author{Kellie Yu Hui Sim}
\email{kellie_sim@mymail.sutd.edu.sg}
\orcid{0009-0005-6451-7089}
\affiliation{
  \institution{Singapore University of Technology and Design}
  \city{Singapore}
  \country{Singapore}
}
\authornote{Corresponding author. Provocation accepted to the CHI 2026 Workshop on Toward Relationship-Centered Care with AI: Designing for Human Connections in Healthcare.}

\author{Kenny Tsu Wei Choo}
\email{kenny_choo@sutd.edu.sg}
\orcid{0000-0003-3845-9143}
\affiliation{
  \institution{Singapore University of Technology and Design}
  \city{Singapore}
  \country{Singapore}
}

\renewcommand{\shortauthors}{Sim and Choo}

\begin{abstract}
    Relationship-centred care (RCC) recognises that healthcare quality depends not only on outcomes, but on how voice, responsibility, and emotional labour are negotiated among patients, caregivers, and providers. 
As AI systems enter sensitive care contexts, they introduce a new participant into these negotiations. 
Drawing on empirical work in Advance Care Planning (ACP) and peer support, we argue that AI's primary impact in high-subjectivity domains is not optimisation but redistribution: it reorganises who speaks, who decides, and who bears moral responsibility. 
Across both settings, participants were less concerned with technical accuracy than with relational consequences: whether AI would appropriately represent their decision, reduce burden, or blur accountability, scaffold connection, or subtly displace it. 
We identify three relational dimensions: authority, temporality, and visibility, through which AI reshapes care relationships, and propose design provocations centred on relational legibility, bounded agency, responsibility traceability, and non-substitutive scaffolding.
\end{abstract}

\maketitle

\section{Background: Relational Care and AI}
In high-subjectivity healthcare contexts, decisions are rarely reducible to objective correctness.
They emerge through negotiation among patients, caregivers, and providers, shaped by trust and shared moral responsibility. 
RCC foregrounds these interdependencies, recognising that care quality depends not only on outcomes but on how voice and accountability are distributed within relationships.
As AI becomes embedded in healthcare, it becomes a new participant in these negotiations.
Depending on its positioning, AI can amplify, mediate, or override human perspectives~\cite{zhangExploringCollaborationPatterns2025}, reshaping how agency and responsibility circulate.

Drawing on empirical work in ACP and peer support, we argue that AI’s primary impact in these domains is redistribution: reorganising who speaks, who decides, and who bears responsibility.
Across both contexts, participants focused less on technical accuracy than on whether AI would represent their decision, clarify accountability, or subtly displace human connection.
In contexts where decisions depend on interpretation, trust, and moral negotiation, the relational positioning of AI may matter more than its computational capability. 
Designing AI that strengthens RCC, therefore, requires examining not only what AI does, but how it is situated within and reshapes human relationships.

\section{Redistribution in Sensitive Contexts}
\subsection{Advance Care Planning: Representation and Surrogate Responsibility}
ACP enables individuals to articulate preferences in anticipation of decisional incapacity~\cite{agarwalAdvanceCarePlanning2018}. 
These preferences are later interpreted by surrogates and clinicians in emotionally charged contexts, making the process inherently relational.

Recent advances in LLMs have enabled conversational systems capable of modelling values, maintaining memory, and generating context-sensitive responses~\cite{songTypingCureExperiences2025, jinDontKnowWhy2025, fooBenefitsRisksLLMs2025, xuGoalsActionsDesigning2025}.
Through persistent representations of user preferences and intentions, such systems have been proposed as forms of persona persistence or generative advocacy~\cite{morrisGenerativeGhostsAnticipating2025}.
In ACP contexts, this raises the possibility of AI agents that represent an individual's values across time.
In prototypes simulating such agents~\cite{simWordsDescribeWhat2026}, participants evaluated them relationally rather than technically.
They questioned whether the system would make their decision or the right decision, highlighting concerns about authorship and moral ownership.

Systems that appeared to reinterpret preferences risked displacing surrogate judgement or obscuring responsibility. 
While stabilising preferences may protect future voice, unclear authority can complicate accountability. 
When carefully bounded, such stabilisation can positively redistribute responsibility by easing surrogate burden and reducing interpretive uncertainty at moments of crisis.
However, when authority is ambiguous, the same redistribution risks obscuring who bears moral and decisional ownership.
In ACP, legitimacy therefore hinges less on predictive competence than on preserving clarity around who ultimately decides.

\subsection{Peer Support and Mental Health: Authenticity and Emotional Labour}
In mental health contexts, alliance and rapport are foundational to effective support~\cite{wampoldAllianceMentalHealth2023}.
Peer support, often grounded in mutual lived experience and relational accessibility~\cite{yeoDigitalPeerSupport2023, shalabyPeerSupportMental2020, solomonPeerSupportPeer2004}, has increasingly been positioned as a scalable response to gaps in formal care, particularly in digitally mediated and community-based settings.
Alongside this expansion, AI systems have been proposed as tools to extend capacity, scaffold conversations, and standardise quality~\cite{hsuHelpingHelperSupporting2025, sharmaHumanAICollaboration2023}.

Drawing on our empirical work involving AI-generated suggestions in peer supporter-client conversations~\cite{simThisReallyHuman2025}, evaluations centred on relational consequence rather than phrasing quality. 
While AI could reduce cognitive strain and offer structure, overly directive suggestions risked shifting responsibility and diluting visible human presence. 
In its more supportive configurations, AI redistributed labour in ways participants welcomed, by offering prompts, grounding, or momentary relief, without undermining relational accountability~\cite{simThisReallyHuman2025}.
Tension arose when this redistribution became opaque or directive, rather than assistive.

Across settings, concerns centred less on phrasing quality than on shifts in rapport, responsibility, and the meaning of care. 
In peer support, legitimacy hinges on preserving authenticity and visible human agency.
The value of AI, hence, depends on whether it scaffolds relational presence or subtly reshapes it.

\section{Relational Dimensions of AI in Care}
Across contexts, AI's impact depended less on the task domain than on its position within the interaction. 
We identify three relational dimensions through which AI reorganises care relationships: authority, temporality, and visibility.

\subsection{Authority: Who Holds Decision Control?}
The introduction of AI shifts perceptions of decision authority, even when formal control remains human. 
Prior work on human-AI collaboration describes agency in terms of locus (who holds primary control), dynamics (how control shifts over time), and granularity (the level at which authority is exercised)~\cite{zhangExploringCollaborationPatterns2025}.

In our empirical work on ACP, participants were wary of systems that appeared to refine or reinterpret their stated preferences~\cite{simWordsDescribeWhat2026}.
Even without explicit autonomy, an AI agent that seemed to "improve" or reinterpret decisions risked being perceived as an authority rather than a representative.
Similarly, in peer support, suggestion systems that felt prescriptive or overly confident were described as subtly shifting responsibility away from the supporter.

Authority often emerged through perception rather than mandate.
The relational risk lies less in explicit automation than in subtle shifts in perceived interpretive legitimacy.
Anthropomorphic cues, fluent language, and emotionally resonant responses can increase empathy and trust toward conversational agents~\cite{liExploringEffectsChatbot2025}.
In sensitive care contexts, such features may inadvertently encourage deference.

\subsection{Temporality: When and For Whom Does AI Speak?}
Temporal horizon significantly shaped how participants evaluated AI's relational legitimacy.
In ACP, decisions articulated in the present are enacted in a future marked by vulnerability and potential conflict. Here, stability and constraint were valued. 
Participants preferred bounded systems with clearly defined scope, minimising the risk that future reinterpretation would complicate surrogate decision-making.
By contrast, peer support involves real-time, emotionally dynamic interactions, with fluidity, pacing, and responsiveness being central.
Systems that interrupted conversational flow or felt rigid were perceived as undermining rapport.
We posit that AI's relational legitimacy is temporally situated: stability may protect voice in ACP, while flexibility preserves authenticity in real-time support.

\subsection{Visibility: How Present Should AI Be?}
AI systems can operate as background scaffolds, visible collaborators, or perceived authorities.
When positioned as background assistants, offering optional suggestions or structuring reflection, AI preserved human-centric control. 
The initiative remained reactive, activated only by explicit invocation.
This configuration was often perceived as supportive rather than intrusive.
When AI became more visible in interactions, relational configurations could shift toward shared agency~\cite{zhangExploringCollaborationPatterns2025}.
Iterative feedback loops may support engagement, yet increased visibility also alters negotiation dynamics.

Importantly, perceptions of authority may arise even without formal delegation.
In high-subjectivity contexts, over-visibility risks repositioning AI from scaffold to quasi-participant. 
RCC depends not only on what AI does, but on how prominently it occupies the relational space.

\section{Provocations for Relationship-Centred AI}
Across contexts, a consistent pattern emerges: AI in healthcare does not simply optimise tasks but reorganises relationships.
In high-subjectivity domains such as ACP and peer support, these redistributions of voice, authority, and responsibility may have consequences that exceed measurable performance gains.
We offer four provocations to guide the design of AI systems aligned with RCC.

\subsection{Relational Legibility Over Model Explainability}
Explainable AI has largely focused on making outputs interpretable and technically transparent~\cite{phillipsFourPrinciplesExplainable2021}. In relational care contexts, however, the more pressing question may be not \emph{why} a system produced a response, but \emph{what position it occupies} within the care relationship.
Designing for relational legibility requires making scope, authority, and limits explicit.
Is the AI advisory, collaborative, or representative? 
Does it refine preferences, summarise them, or merely surface them?
Without explicit signalling of role and constraint, AI risks being perceived as a neutral arbiter or institutional authority, subtly reshaping power dynamics.
Legibility here concerns relational positioning, not only algorithmic reasoning.

\subsection{Bounded Agency Over Immersive Authority}
In morally subjective decisions, such as end-of-life planning or emotionally vulnerable peer conversations, full automation may be neither desirable nor ethically coherent~\cite{shneidermanHumanCenteredAI2022, candrianRiseMachinesDelegating2022}.
Participants across contexts expressed preferences for systems that were clearly bounded in scope and constrained in autonomy~\cite{simWordsDescribeWhat2026, simThisReallyHuman2025}.
Such bounded agency might preserve human interpretive authority, appearing in ACP as resisting systems that could reinterpret preferences beyond their original intent and as avoiding suggestions that seemed prescriptive or emotionally directive in peer support settings. 
Designing for RCC may therefore require resisting immersive or anthropomorphic configurations that encourage deference~\cite{liExploringEffectsChatbot2025}, particularly when relational hierarchies are already fragile.

\subsection{Responsibility Traceability}
When AI enters care negotiations, lines of moral responsibility can blur. 
If AI influences a decision, who is accountable for its consequences: the patient, the surrogate, the supporter, the clinician, or the system designer?
Designing for responsibility traceability involves ensuring that accountability remains legible even when AI contributes to interpretation or articulation. 
Advisory-only configurations, human-in-the-loop escalation pathways, and explicit attribution of decision authority can help preserve moral clarity. 
In sensitive contexts, preserving traceability may matter more than increasing autonomy.

\subsection{Non-Substitutive Scaffolding}
Care relationships are sustained through emotional presence, negotiation, and shared moral labour. 
AI may scaffold articulation and reflection, but it should not substitute the relational work that defines care quality.
Optional suggestions, transparent rationales, and hybrid oversight structures can augment human capacity without displacing human agency. 
In RCC, relational labour is not inefficiency to be optimised away; it is constitutive of care itself. Designing AI that respects this distinction requires treating emotional labour not as overhead, but as core infrastructure.

These provocations invite further discussion:
\begin{itemize}
    \item When does AI amplify voice, and when does it displace it?
    \item Who absorbs emotional and moral labour when AI scales care?
    \item Can AI stabilise preferences across time without destabilising accountability?
    \item What design cues inadvertently grant AI relational authority?
\end{itemize}

\section{Conclusion}
As AI becomes increasingly integrated into healthcare, its most consequential effects may be relational rather than computational. 
In high-subjectivity contexts such as ACP and peer support, AI reorganises who speaks, who decides, and who bears responsibility. 
Designing for RCC, therefore, requires attention not only to performance metrics but to authority, temporality, and visibility within human negotiations.
The challenge is not whether AI can support care, but whether it can do so without obscuring accountability or displacing relational labour.

\bibliographystyle{ACM-Reference-Format} 
\bibliography{main}

\end{document}